# Tracking mm-Wave Channel Dynamics: Fast Beam Training Strategies under Mobility

Joan Palacios[1, 2], Danilo De Donno[1], and Joerg Widmer[1]

[1]IMDEA Networks Institute, Madrid, Spain
[2]Universidad Carlos III de Madrid, Madrid, Spain
E-mail: {firstname.lastname}@imdea.org

*Abstract*—In order to cope with the severe path loss, millimeter-wave (mm-wave) systems exploit highly directional communication. As a consequence, even a slight beam misalignment between two communicating devices (for example, due to mobility) can generate a significant signal drop. This leads to frequent invocations of time-consuming mechanisms for beam re-alignment, which deteriorate system performance. In this paper, we propose smart beam training and tracking strategies for fast mm-wave link establishment and maintenance under node mobility. We leverage the ability of hybrid analog-digital transceivers to collect channel information from multiple spatial directions simultaneously and formulate a probabilistic optimization problem to model the temporal evolution of the mm-wave channel under mobility. In addition, we present for the first time a beam tracking algorithm that extracts information needed to update the steering directions directly from data packets, without the need for spatial scanning during the ongoing data transmission. Simulation results, obtained by a custom simulator based on ray tracing, demonstrate the ability of our beam training/tracking strategies to keep the communication rate only 10% below the optimal bound. Compared to the state of the art, our approach provides a 40% to 150% rate increase, yet requires lower complexity hardware.

## I. Introduction

The fifth generation of mobile communications (5G) is envisaged to deliver multi-Gbps wireless connectivity and to enable a plethora of new applications. It is well established that achieving extremely high data rates is impractical with currently available 4G systems due to the heavily congested and fragmented spectrum below 6 GHz. In view of this, the large amount of unoccupied spectrum in the millimeter wave (mm-wave) bands above 6 GHz becomes very appealing [1].

Communications at mm-wave frequencies are challenging since the channel suffers from severe path loss, atmospheric absorption, human blockage, and other environmental obstructions [2]. The short wavelength of the mm-waves allows beamforming arrays with many antennas to be implemented in a small form factor, thus providing sufficient link margin. On the other hand, highly directional communications complicate the link establishment and maintenance between an Access Point (AP) and a User Equipment (UE). In fact, AP and UE must perform a time-consuming beam training procedure in order to determine the best directions of transmission/reception, which incurs significant overhead (and waste of network resources). The problem is exacerbated in scenarios with mobility, since even a slight beam misalignment or environmental changes, such as link blockage, device rotation, etc., can cause considerable signal drop. To sum up, fast and efficient beam training/tracking strategies are of paramount importance to maintain seamless connectivity in a mm-wave network with node mobility.

The design space of beam search proposals in the literature can be divided into three main categories: (1) sequential scanning strategies [3], [4]; (2) adaptive algorithms employing antenna patterns with configurable beamwidth [5]–[8]; (3) parallel beam search with simultaneous, multi-direction scanning [9], [10]. The vast majority of these works concentrates, however, on static networks without investigating the impact of the training latency on the overall Quality of Service (QoS) of realistic networks with mobility. Within the state-of-the-art solutions on this subject, a further subdivision can be made on the basis of the employed mm-wave transceivers. Since traditional multiple-input multiple-output (MIMO) digital beamforming (DBF) is, at present, impractical at mm-wave frequencies because of cost and power consumption constraints, analog beamforming (ABF) and hybrid analog-digital beamforming (HBF) represent the only feasible solutions. Using ABF [3], [4] provides poor performance for two main reasons. First, the constant amplitude and the low phase resolution constraints of the mm-wave RF phase shifters [11] give rise to antenna sectors with high sidelobes and reduced flatness, leading to imprecise beam training. Second, the use of a single RF chain allows for only one communication beam, thus resulting in reduced throughput and high-overhead beam search. In HBF architectures [5]–[10], the precoding/combining operations are divided between the analog and digital domains, while using much fewer RF chains than antenna elements. The availability of multiple RF chains enables parallel, multi-stream processing and simultaneous multi-direction scanning.

In this paper, we consider a scenario consisting of a fixed AP and a mobile UE, both equipped with a low-complexity mm-wave HBF transceiver and communicating with directional antenna patterns. Our overall objective is to maximize the communication rate over time. To this end, we propose two strategies, a deterministic one for beam training and a probabilistic one for beam tracking, to rapidly estimate multiple, suitable directions of communication between AP and UE. Here, *beam training* is a beam search mechanism without any prior knowledge that explores the entire azimuthal domain and that is carried out both in the initial access phase and, periodically, during the AP-UE communication. *Beam tracking*, instead, is an ongoing estimation that, starting from the current steering directions, probabilistically infers how they evolve due to node mobility. The main contributions of the paper are as follows:

- We design a two-stage beam training protocol that approaches the performance of an exhaustive search over all possible beam directions, but has very low latency and uses implicit feedback (i.e., it does not require a dedicated feedback channel). The key aspect of our beam search strategy is a particular HBF combiner matrix which takes a reduced number of sequential, multi-stream signal measurements to cover all the possible combinations of antenna weights.
- We propose, to the authors' knowledge for the first time in the mm-wave HBF context, a beam tracking algorithm that is able to track the mm-wave channel dynamics without any training slots, but simply using known portions of the data packet (e.g., the preamble). To this end, we formulate a probabilistic optimization problem, solved by gradient descent, whose objective function is designed so as to model the temporal evolution of channel paths due to device movements. Note that this problem is quite different from the problem of MIMO channel estimation using known pilot symbols.
- We develop a simulation framework to assess the performance of the proposed beam training/tracking strategies and compare them against existing approaches in the literature. Specifically, we propose and implement a fast protocol for link establishment and maintenance under user mobility which dynamically switches between beam training and beam tracking based on the real-time QoS. Our simulator integrates a ray-tracing tool to accurately model the time-varying mm-wave channel, taking into account blockage, ray clustering, and mobility effects, and guaranteeing spatial consistency over time.

Numerical experiments show that the performance provided by our solution is very close to the optimal "oracle-based" algorithm. Furthermore, the high accuracy and reduced latency overhead characterizing our beam training/tracking strategies result in a significant rate increase over state-of-the-art solutions which in addition require higher complexity hardware. Compared to ABF solutions which share the disadvantage of converging towards only one communication beam, our approach based on HBF is capable of achieving multiplexing gains by simultaneously transmitting multiple parallel data streams over different paths.

We use the following notation in the paper. $\mathbf{A}$ is a matrix, $\mathbf{a}$ is a vector, and $\mathcal{A}$ denotes a set. $\|\mathbf{a}\|_2$ is the Euclidean norm of $\mathbf{a}$, while $\|\mathbf{A}\|_F$, $|\mathbf{A}|$, $\mathbf{A}^T$, $\mathbf{A}^H$, and $\mathbf{A}^{-1}$ denote the Frobenius norm, determinant, transpose, Hermitian, and inverse of $\mathbf{A}$, respectively. $[\mathbf{A}]_{\mathcal{B},:}$ ($[\mathbf{A}]_{:,\mathcal{B}}$) are the rows (columns) of the matrix $\mathbf{A}$ indexed by the set $\mathcal{B}$, and $\mathbf{I}$ is the identity matrix. $\mathbb{E}[\cdot]$ denotes the expectation operator and $\lceil \cdot \rceil$ the ceiling function.

## II. Related work

Most of the literature on mm-wave beam search focuses on static scenarios without user mobility [4]–[8], [10]. Such an assumption may lead to wrong conclusions about the actual performance of the algorithms in real networks. A comparative analysis of initial access techniques in mm-wave networks is presented in [4], where performance metrics such as detection probability and delay are analyzed. The problem of tracking the AP-UE beams to handle the channel dynamics is left as future work. The design of HBF codebooks relying on beamforming vectors with different beamwidths is presented in [5]–[8], where it is assumed that phase shifters with a large number of quantization bits are available at mm-wave frequencies. However, the design of high-resolution mm-wave phase shifters is extremely challenging [11]. Finally, the simultaneous reception from multiple beams to accelerate the beam search is exploited in [10].

To the authors' knowledge, only very few works in the literature address the problem of fast beam search in realistic, dynamic scenarios with node mobility. A smart beam steering algorithm for 60 GHz link re-establishment under user mobility is presented in [3]. The algorithm uses knowledge of previous feasible sector pairs to narrow the sector search space, thereby reducing the associated overhead. Numerical results show that the proposed strategy is very effective, but still incurs non-negligible latency in complex scenarios with significant blockage. A temporal channel evolution model for non line-of-sight (NLOS) mm-wave scenarios is presented in [9]. HBF at both the AP and UE is considered and a beam tracking technique based on sequentially updating the precoder and combiner is developed. However, in [9], the angle of arrival (AoA) and angle of departure (AoD) deviations due to mobility are modeled as very small uniform random variables, which are not appropriate to characterize actual mobility or significant, sudden changes in the channel due to obstacles. In [12], a linear dynamic system model to analyze the occurring errors due to link blockage and device movement is proposed. Based on the model, the authors propose two probing protocols that are effective in identifying the beam errors. However, no beam training/tracking strategy is implemented in order to find alternative antenna sector pairs once the beam errors are identified. Finally, it is worth highlighting that none of the above-mentioned works [3], [9], [12] analyzes the impact of the beam search accuracy and overhead on the evolution over time of the achievable rate under mobility.

## III. Motivation and system model

The use of highly directional antennas with very narrow beams at both the AP and UE complicates the mm-wave link establishment and maintenance. As for the link establishment, the 60-GHz IEEE 802.11ad standard [13] implements a time-consuming beam training procedure based on an exhaustive search to find the most suitable directions of transmission and reception. Once a connection is established, the link quality degradation due to user mobility is handled through beam refinement procedures that search around the previous sector pair in order to determine a new combination of beams with improved link quality. However, in large and crowded scenarios with mobility, such procedures may fail to cope with high channel dynamics, which would necessitate fast mechanisms to scan a large angular domain (instead of just adjacent directions) to find alternative communication links. In case simply probing adjacent beams is unsuccessful, a new exhaustive beam search procedure has to be performed. This leads to a high latency which deteriorates the overall system performance. Motivated by this challenging problem, we propose two smart and efficient strategies, one for beam training and one for beam tracking, to accelerate the link establishment and maintenance between mm-wave devices in mobility scenarios.

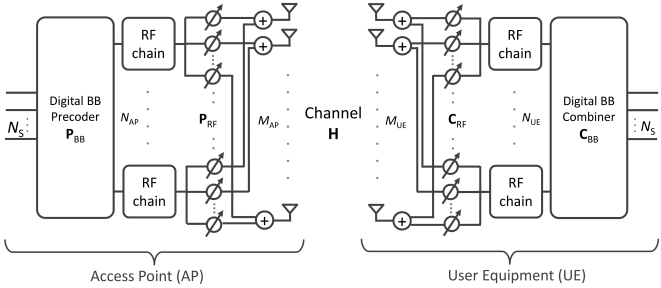

Fig. 1. Block diagram of the AP-UE mm-wave transceiver architecture implementing HBF.

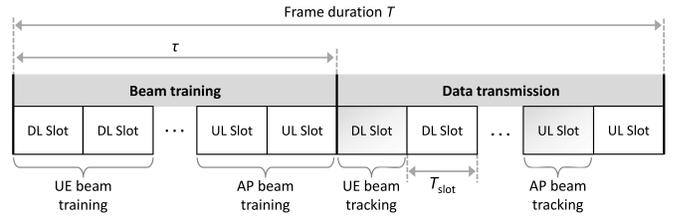

Fig. 2. Frame structure encompassing beam training/tracking and data transmission. Data slots can be indifferently either downlink (DL) or uplink (UL) slots.

We consider a mm-wave system with one fixed AP and one moving UE, both featuring the same HBF architecture considered in [5], [6], [8]–[10] and depicted in Fig. 1. The AP is equipped with a uniform linear array (ULA) of $M_{\text{AP}}$ isotropic radiators connected to $N_{\text{AP}}$ RF transceiver chains through a network of analog/RF phase shifters. The number of antennas and RF transceiver chains at the UE side is $M_{\text{UE}}$ and $N_{\text{UE}}$ respectively. The HBF transceiver configuration allows AP and UE to communicate via $N_{\text{S}}$ data streams, with $N_{\text{S}} \leq \min(N_{\text{AP}}, N_{\text{UE}})$. To this end, the AP applies an $N_{\text{AP}} \times N_{\text{S}}$ digital baseband (BB) precoder $\mathbf{P}_{\text{BB}}$ followed by an $M_{\text{AP}} \times N_{\text{AP}}$ RF precoder, $\mathbf{P}_{\text{RF}}$, to the symbol sequence to be transmitted. The transmit power constraint is ensured by imposing $\|[\mathbf{P}_{\text{RF}}\mathbf{p}_{\text{BB}}]_{:,i}\|_2^2 = 1$, for $i = 1, 2, ..., N_{\text{S}}$. The final AP precoder is then given by the $M_{\text{AP}} \times N_{\text{S}}$ matrix $\mathbf{P} = \mathbf{P}_{\text{RF}}\mathbf{P}_{\text{BB}}$. The transmitted signal passes through the $M_{\text{UE}} \times M_{\text{AP}}$ channel matrix $\mathbf{H}$ and impinges on the UE antennas together with white noise. Since the UE also implements HBF, it is able to concurrently receive $N_{\text{S}}$ streams of data. To do that, it applies a $M_{\text{UE}} \times N_{\text{UE}}$ RF combiner $\mathbf{C}_{\text{RF}}$ followed by a $N_{\text{UE}} \times N_{\text{S}}$ digital baseband combiner $\mathbf{C}_{\text{BB}}$. The final UE combiner is given by the $M_{\text{UE}} \times N_{\text{S}}$ matrix $\mathbf{C} = \mathbf{C}_{\text{RF}}\mathbf{C}_{\text{BB}}$.

We assume that AP and UE communicate using the frame structure in Fig. 2. Two different types of frames can be allocated: (i) beam training frames, which contain both training and data transmission phases, and (ii) pure data frames. In the initial access procedure, the allocation of a training frame is mandatory, since AP and UE need to determine suitable initial directions of transmission. Once the initial access is accomplished, pure data frames with directional antenna patterns at both the AP and UE are used. As explained later in §V, the beam tracking can be performed with pure data frames, i.e., using known portions of just two data slots (one for UE beam tracking and one for AP beam tracking) without requiring any dedicated training slots. The allocation of a training frame to perform a new full beam search from scratch can be triggered periodically or when the link quality falls below a certain threshold. Based on the work in [14], [15], we assume frames of duration $T$=10 ms, each divided into 100 slots of duration $T_{\text{slot}}$=100 $\mu$s, a sufficiently small value to ensure channel coherence at mm-wave frequencies.

As experimentally demonstrated in [16], the mm-wave channel between AP and UE is composed of "ray clusters", each cluster carrying a fraction of the total power. Defining $T_{\text{slot}}$=100 $\mu$s as the time granularity of our system, we can express the $M_{\text{UE}} \times M_{\text{AP}}$ channel matrix at each time slot as:

$$\mathbf{H} = \sqrt{\frac{M_{\text{AP}}M_{\text{UE}}}{L}} \sum_{k=1}^{K}\sum_{\ell=1}^{L} \alpha_{k\ell}\mathbf{a}_{\text{UE}}(\theta_{k\ell})\mathbf{a}_{\text{AP}}^{H}(\phi_{k\ell}) \qquad (1)$$

where $K$ is the number of clusters, $L$ is the number of sub-paths per cluster, $\mathbf{a}_{\text{UE(AP)}}(\cdot)$ is the ULA response vector at the UE (AP) whose expression can be found in [8, Eq. (3)], and $\alpha_{k\ell}$ is the complex gain on the $\ell$-th sub-path of the $k^{\text{th}}$ cluster, which includes path loss, Doppler shift, and delay spread effects. The variables $\theta_{k\ell} \in [0, 2\pi]$ and $\phi_{k\ell} \in [0, 2\pi]$ are the $\ell^{\text{th}}$ AoD/AoA of the $k^{\text{th}}$ cluster at the UE and AP respectively. In this work, we assume channel reciprocity [5], [6], that is, the AP AoDs in the downlink correspond to the AP AoAs in the uplink. The same applies to the UE as well. Note that, in order to simplify the notation, we consider the AP and UE implementing horizontal (2-D) beamforming only, which implies that all scattering happens in the azimuthal domain. Extension to planar antenna arrays and, therefore, to 3-D beamforming is straightforward.

IV. PSEUDO-EXHAUSTIVE BEAM TRAINING (PE-TRAIN)

The use of directional antennas for mm-wave communication requires that AP and UE find suitable directions of transmission, both in the initial access phase and, periodically, during the communication. As illustrated is Fig. 2, if no feedback channel is available, two separate stages are required in the beam training phase, namely UE beam training (using downlink training sequences from the AP) and AP beam training (using uplink training sequences from the UE). In the following, we propose a pseudo-exhaustive beam training (PE-Train) protocol which is able to search over all possible beam directions with very low latency overhead. It uses omnidirectional transmission at the AP for UE beam training, and simultaneous, multi-stream transmission over the best estimated directions at the UE for AP beam training.

## A. Stage I: UE beam training

We consider a mm-wave AP with the HBF architecture in Fig. 1 performing omnidirectional transmission of a training sequence $s[t]$, for $t = 1, 2, ..., T_s$, in a reciprocal channel. Arranging $s[t]$ into the $1 \times T_s$ row vector $\mathbf{s}$, the $M_{\text{UE}} \times T_s$ discrete-time signal $\mathbf{R}$ impinging on the UE antennas becomes:

$$\mathbf{R} = \sqrt{P_t}\mathbf{H}\mathbf{p}_\text{o}\mathbf{s} + \mathbf{N} \quad (2)$$

where $P_t$ is the transmit power, $\mathbf{H}$ is the channel matrix, $\mathbf{p}_\text{o} = [1, 0, 0, ..., 0]^T$ is the $M_{\text{AP}} \times 1$ omnidirectional precoding vector used at the AP, and $\mathbf{N}$ is a $M_{\text{UE}} \times T_s$ matrix with independent, Gaussian-distributed complex noise with mean zero and variance $\sigma^2$. The lack of a dedicated RF chain for each antenna makes it impossible for a HBF transceiver to directly access $\mathbf{R}$. In fact, $\mathbf{R}$ is inevitably processed by a hybrid combiner which compresses it into a reduced dimension space, with consequent loss of information. We tackle this problem with the following strategy. First, we design an easily invertible, orthogonal $M_{\text{UE}} \times M_{\text{UE}}$ matrix $\mathbf{W}$ (e.g., a Hadamard matrix) representing a basis for the full space of antenna configurations. Then, since we cannot directly apply $\mathbf{W}$ to $\mathbf{R}$ because of the HBF limitations, we perform multiple, consecutive measurements, each time using as hybrid combiner a different sub-matrix of $\mathbf{W}$. Thanks to the properties of $\mathbf{W}$, we can reconstruct, at the end of the procedure, an estimated version of $\mathbf{R}$ and process it through a spatial filter matrix to derive the received signal power from each angular direction. Specifically, we build $\mathbf{W}$ such that the elements of $\sqrt{N_{\text{UE}}}\mathbf{W}$ belong to the feasible set of phase-shifter weights. For example, in the case of 2-bit phase shifters, the elements of $\sqrt{N_{\text{UE}}}\mathbf{W}$ can assume only four values, namely $\pm 1$ and $\pm j$. Then, we divide $\mathbf{W}$ into $N_{\text{W}} = \lceil M_{\text{UE}}/N_{\text{UE}} \rceil$ sub-matrices with dimensions $M_{\text{UE}} \times N_{\text{UE}}$, i.e., $\mathbf{W} = [\mathbf{W}_1, \mathbf{W}_2, ..., \mathbf{W}_{N_{\text{W}}}]$. For each $\mathbf{W}_i$, with $i = 1, 2, ..., N_{\text{W}}$, we build the RF combiner $\mathbf{C}_{\text{RF},i} = \sqrt{N_{\text{UE}}}\mathbf{W}_i$ and the baseband combiner $\mathbf{C}_{\text{BB},i} = \mathbf{I}_{N_{\text{UE}}}/\sqrt{N_{\text{UE}}}$, where $\mathbf{I}_{N_{\text{UE}}}$ is the $N_{\text{UE}} \times N_{\text{UE}}$ identity matrix. The overall hybrid combiner $\mathbf{C}_i = \mathbf{C}_{\text{RF},i}\mathbf{C}_{\text{BB},i} = \mathbf{W}_i$ is then applied to $\mathbf{R}$ at $N_{\text{W}}$ successive instants in order to extract the following signal measurements:

$$\mathbf{Y}_i = \mathbf{W}_i^H(\sqrt{P_t}\mathbf{H}\mathbf{p}_\text{o}\mathbf{s} + \mathbf{N}), \ i = 1, 2, ..., N_{\text{W}} \quad (3)$$

which can be concatenated as $\mathbf{Y} = [\mathbf{Y}_1^T, \mathbf{Y}_2^T, ..., \mathbf{Y}_{N_{\text{W}}}^T]^T$. Pure channel information can be extracted by removing the contribution of the training sequence, which is supposed to be known at the UE. To do this, we compute $\hat{\mathbf{Y}} = \mathbf{Y}\mathbf{s}^H/T_s$, whose expected value $\mathbb{E}[\hat{\mathbf{Y}}] = \sqrt{P_t}\mathbf{W}^H\mathbf{H}\mathbf{p}_\text{o}$ gives direct access to channel-only information. The UE can now post-process the measurement $\hat{\mathbf{Y}}$ to obtain:

$$\overline{\mathbf{Y}} = \mathbf{A}_{\text{UE}}^H(\mathbf{W}^H)^{-1}\hat{\mathbf{Y}} \quad (4)$$

where $\mathbf{A}_{\text{UE}} = [\mathbf{a}_{\text{UE}}(\theta_1), \mathbf{a}_{\text{UE}}(\theta_2), ..., \mathbf{a}_{\text{UE}}(\theta_N)]$ is a spatial filter matrix and $\theta_i = \frac{2\pi i}{N}$, $i = 1, 2, ..., N$, is a set of $N$ equally spaced discrete angles covering the 360° azimuthal domain. As evident from Eq. 4 and from the expression $\mathbb{E}[\overline{\mathbf{Y}}] = \sqrt{P_t}\mathbf{A}_{\text{UE}}^H\mathbf{H}\mathbf{p}_\text{o}$, the $N \times 1$ vector $|[\overline{\mathbf{Y}}]_i|^2$, for $i = 1, 2, ...N$, contains the expected signal power impinging on the UE from each angular direction. Such information can be directly used by the UE to estimate its $L_{\text{est}}$ most suitable (i.e., the most powerful) directions of transmission/reception $\boldsymbol{\theta} = [\theta_1, ..., \theta_{L_{\text{est}}}]$. Since the product $\mathbf{A}_{\text{UE}}^H(\mathbf{W}^H)^{-1}$ in Eq. 4 can be precomputed and stored in the UE memory, the computational cost to estimate $\overline{\mathbf{Y}}$ is just a matrix-vector multiplication.

## B. Stage II: AP beam training

At the end of Stage I, the UE initiates the AP beam training stage. Specifically, the UE employs the HBF algorithms in [10] (considering 2-bit RF phase shifters) to design a multi-beam/multi-stream precoder $\mathbf{P}$ with the narrowest synthesizable beamwidth. Such a precoder is then used to simultaneously transmit orthogonal training sequences through the set of directions $\boldsymbol{\theta}$ estimated in Stage I. In order to reduce the inter-beam interference, Golay training sequences encoded by orthogonal Walsh spreading codewords are used. In fact, Golay sequences possess very good auto-correlation, which helps protecting the Walsh codes from losing orthogonality due to multipath. Concretely, for each transmit direction $i = 1, 2, ..., L_{\text{est}}$, with $L_{\text{est}} \leq N_{\text{UE}}$, the UE emits a training signal $s_i[t]$, for $t = 1, 2, ..., T_s$. The overall set of transmitted symbols can be arranged into a matrix $\mathbf{S}$, with dimensions $L_{\text{est}} \times T_s$, where the $i$-th row contains the time-domain sequence transmitted over the $i$-th direction. As in Stage I, the AP builds a $M_{\text{AP}} \times M_{\text{AP}}$ matrix $\mathbf{W}$ and configures its hybrid combiner to perform $N_{\text{W}} = \lceil M_{\text{AP}}/N_{\text{AP}} \rceil$ signal measurements at successive instants:

$$\mathbf{Y}_i = \mathbf{W}_i^H(\rho\mathbf{H}\mathbf{P}\mathbf{S} + \mathbf{N}), \ i = 1, 2, ..., N_{\text{W}} \quad (5)$$

where $\rho = \sqrt{P_t/L_{\text{est}}}$ for equally distributed power within the streams. Similar to what is done in Stage I, the AP then concatenates the measurements, estimates $\hat{\mathbf{Y}} = \mathbf{Y}\sqrt{L_{\text{est}}}\mathbf{S}^H/T_s$, and processes it with the spatial filter matrix $\mathbf{A}_{\text{AP}}$ to obtain $\overline{\mathbf{Y}}$. The procedure allows the AP to estimate its $L_{\text{est}}$ most suitable directions of transmission/reception, $\boldsymbol{\phi} = [\phi_1, ..., \phi_{L_{\text{est}}}]$. In order to establish a multi-beam, multi-stream data link between AP and UE after beam training, it is necessary that $L_{\text{est}} \leq \min(N_{\text{AP}}, N_{\text{UE}})$. Note that the computational cost of estimating one rather than more than one suitable directions for communication via PE-Train is the same.

The major algorithmic steps required by the PE-Train procedure are summarized in Algorithm 1 for $M_{\text{AP}}$ and $M_{\text{UE}}$ integer multiples of $N_{\text{AP}}$ and $N_{\text{UE}}$ respectively. The beam training overhead, i.e., the total time required to complete the PE-Train procedure, is given by the sum of Stage I and Stage II times: $\tau = \tau_{\text{PE-Train}} = T_{\text{slot}}(\lceil M_{\text{UE}}/N_{\text{UE}} \rceil + \lceil M_{\text{AP}}/N_{\text{AP}} \rceil)$. This represents a significant speed-up compared to exhaustive beam training, where the time required to complete the beam search is $\tau = \tau_{\text{EXH}} = T_{\text{slot}}N^2$, where $N$ is the same angular resolution of our PE-Train strategy.

**Algorithm 1** PE-Train protocol
---
**Initialization:** Pre-compute and store $\mathbf{A}_{\text{AP}}$ and $\mathbf{A}_{\text{UE}}$
**Stage I:** UE beam training
1: AP in omni mode w/ precoder $\mathbf{p}_o = [1, 0, 0, ..., 0]^T$
2: UE pre-computes and stores its $M_{\text{UE}} \times M_{\text{UE}}$ matrix $\mathbf{W}$
3: $N_W = M_{\text{UE}}/N_{\text{UE}}$
4: **for** $i \leq N_W$ **do**
5: $\quad \mathbf{W}_i = [\mathbf{W}]_{:,(i-1)N_{\text{UE}}+1:iN_{\text{UE}}}$
6: **end for**
7: **for** $i \leq N_W$ **do**
8: $\quad$ UE measures $\mathbf{Y}_i = \mathbf{W}_i^H(\sqrt{P_t}\mathbf{H}\mathbf{p}_o\mathbf{s} + \mathbf{N})$
9: **end for**
10: $\mathbf{Y} = [\mathbf{Y}_1^T, ..., \mathbf{Y}_{N_W}^T]^T$; $\overline{\mathbf{Y}} = \mathbf{A}_{\text{UE}}^H(\mathbf{W}^H)^{-1}\mathbf{Y}\mathbf{s}^H/T_s$
11: **for** $\ell \leq L_{\text{est}}$ **do**
12: $\quad k = \arg\max_j |[\overline{\mathbf{Y}}]_j|^2$; $\theta_\ell = 2\pi k/N$
13: $\quad \overline{\mathbf{Y}} = \overline{\mathbf{Y}} - [\mathbf{A}_{\text{UE}}]_{:,k}^H \mathbf{A}_{\text{UE}}[\overline{\mathbf{Y}}]_k$
14: **end for**
15: **return** $\boldsymbol{\theta} = [\theta_1, ..., \theta_{L_{\text{est}}}]$
**Stage II:** AP beam training
16: UE in multi-beam mode over $\boldsymbol{\theta}$ w/ hybrid precoder $\mathbf{P}$
17: AP pre-computes and stores its $M_{\text{AP}} \times M_{\text{AP}}$ matrix $\mathbf{W}$
18: $N_W = M_{\text{AP}}/N_{\text{AP}}$
19: **for** $i \leq N_W$ **do**
20: $\quad \mathbf{W}_i = [\mathbf{W}]_{:,(i-1)N_{\text{AP}}+1:iN_{\text{AP}}}$
21: **end for**
22: **for** $i \leq N_W$ **do**
23: $\quad$ AP measures $\mathbf{Y}_i = \mathbf{W}_i^H(\sqrt{P_t/L_{\text{est}}}\mathbf{HPS} + \mathbf{N})$
24: **end for**
25: $\mathbf{Y} = [\mathbf{Y}_1^T, ..., \mathbf{Y}_{N_W}^T]^T$; $\overline{\mathbf{Y}} = \mathbf{A}_{\text{AP}}^H(\mathbf{W}^H)^{-1}\mathbf{Y}\sqrt{L_{\text{est}}}\mathbf{S}^H/T_s$
26: **for** $\ell \leq L_{\text{est}}$ **do**
27: $\quad k = \arg\max_j |[\overline{\mathbf{Y}}]_{j,\ell}|^2$; $\phi_\ell = 2\pi k/N$
28: **end for**
29: **return** $\boldsymbol{\phi} = [\phi_1, ..., \phi_{L_{\text{est}}}]$

## V. PROBABILISTIC BEAM TRACKING (P-TRACK)

The PE-Train procedure described in the previous section can be used for initial access beam training, but can also be triggered periodically in order for AP and UE to update their steering directions. However, harsh environments with frequent blockage could lead to an excessive number of beam training requests, which incur significant overhead and, consequently, reduced throughput. For this reason, efficient beam tracking strategies are required in order to rapidly refine the beam directions without resorting to full beam training.

In this section, we propose a probabilistic beam tracking (P-Track) mechanism which is able to track the mm-wave channel dynamics under node mobility (and steer the device beams accordingly) without requiring dedicated training slots. We assume a fixed AP and a moving UE that have just accomplished the PE-Train procedure and are communicating using pure data frames (i.e., without dedicated training slots) and highly directional beam patterns. For the sake of brevity, we consider only the UE beam tracking procedure, i.e., the procedure by which the UE exploits downlink data slots to refine its beam directions. An identical strategy is applied for AP beam tracking using uplink data slots. The P-Track strategy is based on a probabilistic model which does not require the devices to perform any spatial scanning during the ongoing data communication. It is able to track the most dominant directions of the mm-wave channel using just known portions of the data packet, e.g., the preamble. To do that, we require that two conditions are satisfied: (1) the preamble is correctly detected by the UE in at least one downlink slot within the frame; (2) in such a downlink slot, the UE can access the complex output $\mathbf{Y}_{\text{RF}}$ from the RF combiner $\mathbf{C}_{\text{RF}}$ (which can be done by saving the preamble samples right before the baseband combiner $\mathbf{C}_{\text{BB}}$):

$$\mathbf{Y}_{\text{RF}} = \mathbf{C}_{\text{RF}}^H \left( \sqrt{P_t/L_{\text{est}}}\mathbf{HPS} + \mathbf{N} \right) \quad (6)$$

where $P_t$ is the AP transmit power, $L_{\text{est}}$ is the number of parallel data streams transmitted by the AP using the $M_{\text{AP}} \times L_{\text{est}}$ hybrid precoder $\mathbf{P}$, and $\mathbf{S}$ is the $L_{\text{est}} \times T_s$ matrix encompassing the packet preamble transmitted simultaneously by the AP over $L_{\text{est}}$ directions. We assume that AP and UE are communicating using the narrowest beam patterns they are able to synthesize, which are steered towards the spatial directions estimated in the most recent PE-Train/P-Track procedure. In order to reduce the interference among the parallel streams, we assume that, for the preamble, the AP adopts Golay sequences encoded with orthogonal Walsh spreading codewords. The reason behind the choice of using $\mathbf{Y}_{\text{RF}}$ for beam tracking, instead of the signal $\mathbf{Y} = \mathbf{C}_{\text{BB}}^H \mathbf{Y}_{\text{RF}}$ after the baseband precoder, is that the former includes much more information about the channel than the latter which is defined in a lower dimensional space. With $\mathbf{Y}_{\text{RF}}$ it is possible to provide channel information for a wider angular domain compared to the very narrow angular sector covered by the actual data communication beam pattern.

### A. Probabilistic optimization problem

Since the AP is transmitting relevant data to the UE using pure data frames, the UE cannot perform any beam scan, but it must keep its antenna steered towards the directions estimated in the most recent PE-Train/P-Track execution[1]. We propose a probabilistic estimation based on the analysis of the preamble signal $\mathbf{Y}_{\text{RF}}$ received by the UE over the current antenna pattern in a downlink data slot. The UE is moving along a mobility pattern or *route*, so the objective is to update, in real time, its antenna pattern based on the estimation of a new set $\boldsymbol{\theta}^*$ of suitable directions. This can be translated into the problem of finding the $\boldsymbol{\theta}^*$ that maximizes the probability $P(\boldsymbol{\theta}^*|\mathbf{Y}_{\text{RF}})$:

$$\boldsymbol{\theta}^* = \arg\max_{\boldsymbol{\theta}} P(\boldsymbol{\theta}^*|\mathbf{Y}_{\text{RF}}) \qquad (7)$$

which cannot be easily handled because both the prior distribution of the channel and the hybrid precoder $\mathbf{P}$ used by the AP are unknown. Applying the negative logarithm function and the Bayes theorem to the objective function in Eq. 7, and observing that the term $P(\mathbf{Y}_{\text{RF}})$ has no impact on the optimization, we obtain:

$$\boldsymbol{\theta}^* = \arg\min_{\boldsymbol{\theta}} [O_P(\boldsymbol{\theta}^*) + O_Y(\boldsymbol{\theta}^*)] \qquad (8)$$

where we designate $O_P(\boldsymbol{\theta}^*) = -\log[P(\boldsymbol{\theta}^*)]$ as prior objective function and $O_Y(\boldsymbol{\theta}^*) = -\log[P(\mathbf{Y}_{\text{RF}}|\boldsymbol{\theta}^*)]$ as measurement objective function. Intuitively, $O_Y(\boldsymbol{\theta}^*)$ models those changes in the mm-wave channel, due to user mobility, that are directly reflected into the received signal. In contrast, $O_P(\boldsymbol{\theta}^*)$ captures the uncertainties on the previous estimate and propagation phenomena which cannot be inferred from signal measurements.

---
[1]Here, the problem is more complex than channel estimation in MIMO systems. In fact, while MIMO transceivers (based on DBF) are able to instantaneously collect full spatial information about the channel, HBF transceivers would usually need several estimation steps, during which the beams are steered to scan different spatial directions.

As for the prior objective function $O_P(\boldsymbol{\theta}^*)$, a simple approach is to define the prior distribution $P(\boldsymbol{\theta}^*)$ as a set of independent Gaussian distributions for each $\theta_\ell$, $\ell = 1, 2, ..., L_{\text{est}}$, with mean $\overline{\theta}_\ell$ equal to the previous estimated direction and standard deviation $\sigma_{\theta_\ell} = \overline{\sigma}_{\theta_\ell} + f(\text{SNR})$. The term $\overline{\sigma}_{\theta_\ell}$ models the uncertainty of the estimation due to continuous angular variations induced by the UE movements (device rotation and translation), while $f(\text{SNR})$ is a convenient, monotonically decreasing function of the SNR. This latter term models the uncertainty of our previous estimation, which is largely affected by the received signal quality (i.e., the greater the SNR, the smaller the uncertainty). After removing constant terms, the following expression for $O_P(\boldsymbol{\theta})$ can be derived:

$$O_P(\boldsymbol{\theta}^*) = -\sum_{\ell=1}^{L_{\text{est}}} \log P(\boldsymbol{\theta}^*) = \sum_{\ell=1}^{L_{\text{est}}} \frac{(\theta_\ell - \overline{\theta}_\ell)^2}{2\sigma_{\theta_\ell}^2} \qquad (9)$$

In Appendix A, we show that the measurement objective function $O_Y(\boldsymbol{\theta}^*)$ can be expressed as:

$$O_Y(\boldsymbol{\theta}^*) = \frac{\|\mathbf{D}^{-1}\mathbf{V}^H \hat{\mathbf{Y}}_{\text{RF}} - \mathbf{U}^H \mathbf{A}_\theta \mathbf{M}\|_F^2}{2\sigma^2} \qquad (10)$$

where $\mathbf{U}\mathbf{D}\mathbf{V}^H$ is the economic singular value decomposition (SVD) of $\mathbf{C}_{\text{RF}}$, $\hat{\mathbf{Y}}_{\text{RF}} = \mathbf{Y}_{\text{RF}}\sqrt{L_{\text{est}}}\mathbf{S}^H/T_s$ is the redundancy-free preamble signal from the RF precoder, $\sigma^2$ is the noise power, $\mathbf{A}_\theta$ is a $M_{\text{UE}} \times L_{\text{est}}$ matrix such that $[\mathbf{A}_\theta]_{:,\ell} = \mathbf{a}_{\text{UE}}(\theta_\ell)$, and $\mathbf{M} = (\mathbf{A}_\theta^H \mathbf{U}\mathbf{U}^H \mathbf{A}_\theta)^{-1}\mathbf{A}_\theta^H \mathbf{U}\mathbf{D}^{-1}\mathbf{V}^H \hat{\mathbf{Y}}_{\text{RF}}$ is a $L_{\text{est}} \times L_{\text{est}}$ matrix whose derivation can be found in Appendix A.

### B. Problem solution

The problem of estimating $\boldsymbol{\theta}^*$ can be transformed into the problem of minimizing the overall objective function $O(\boldsymbol{\theta}^*) = O_P(\boldsymbol{\theta}^*) + O_Y(\boldsymbol{\theta}^*)$, which can be verified to be non-convex even for $L_{\text{est}} = 1$. The minimization strategy we propose is as follows. Two suitable initial guesses of $\boldsymbol{\theta}^*$, one for $O_P(\boldsymbol{\theta}^*)$ and one for $O_Y(\boldsymbol{\theta}^*)$, are selected. A good initial guess is $\overline{\boldsymbol{\theta}}$, since it minimizes $O_P(\boldsymbol{\theta}^*)$. A second good starting point representing a suitable guess for $O_Y(\boldsymbol{\theta}^*)$ is the set of directions with maximum received signal power which can be computed as explained in Appendix B. Starting from these two initial solutions, since we do not know a priori if $O_Y(\boldsymbol{\theta}^*)$ has more weight than $O_P(\boldsymbol{\theta}^*)$ on the total objective function $O(\boldsymbol{\theta}^*)$, we compute a few steps of gradient descent for both guesses, select the best solution, and proceed with a more accurate gradient descent to refine the estimation. Closed-form expressions for the gradient of $O_P(\boldsymbol{\theta}^*)$ and $O_Y(\boldsymbol{\theta}^*)$ are derived in Appendix B.

## VI. NUMERICAL EVALUATION

In this section, we carry out numerical experiments to assess the performance of the PE-Train and P-Track strategies. We first describe the simulator we developed to evaluate mm-wave indoor scenarios with mobility. Then, we describe the simulation scenario, based on which we conduct a numerical evaluation to compare the performance of our strategies against existing beam search approaches in the literature.

### A. Simulator overview

The main functional blocks of our mm-wave simulator are outlined in Fig. 3. The simulator allows to draw any UE route in a given scenario via a graphical user interface. Based on the route length $L_{\text{route}}$ and the selected UE speed $v$, the number of frames $\nu = \lceil L_{\text{route}}/(vT) \rceil$ required by the simulation are computed. Implicitly, this creates a direct correspondence between the UE position on the route and the current time slot. We exploit such correspondence to compute, at each time slot, the channel matrix $\mathbf{H}$ in Eq. 1 through a proprietary ray tracing program written in C (see the next subsection for details). The PE-Train and P-Track strategies are implemented in Matlab and validated as follows (see Fig. 4 for an example use case). In the UE starting position, the simulator allocates a training frame to perform the initial access PE-Train procedure and establish a multi-beam/multi-stream directional link between AP and UE. Specifically, AP and UE employ the HBF algorithms in [10] (with 2-bit phase shifters) to design the narrowest synthesizable beams, steered toward the $L_{\text{est}}$ estimated directions, to be used concurrently during the data transmission phase. From here on, as long as some QoS or timing conditions (defined later) are satisfied, pure data frames are sent. In each frame, a downlink data slot and an uplink data slot are used by UE and AP respectively to perform the P-Track estimation and update, accordingly, their steering directions. If the QoS and timing conditions are not satisfied, a training frame is allocated in order to perform a thorough and accurate beam search via PE-Train. Concretely, for each frame, we compute two types of sum-rate capacity (hereafter referred to as simply rate):

- The actual rate $R_D$ within the current frame is obtained by computing the rate in each data transmission slot and averaging it *over the number of data slots in the frame*.

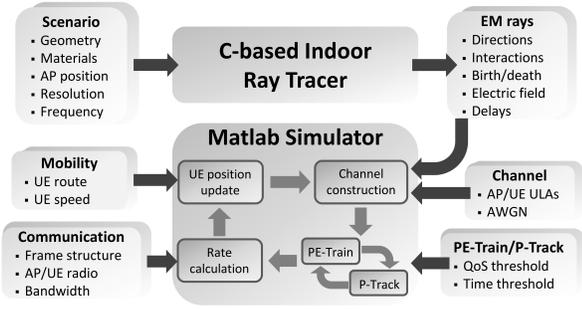

Fig. 3. Overview of the mm-wave indoor simulator used to assess the performance of beam training and tracking strategies under node mobility.

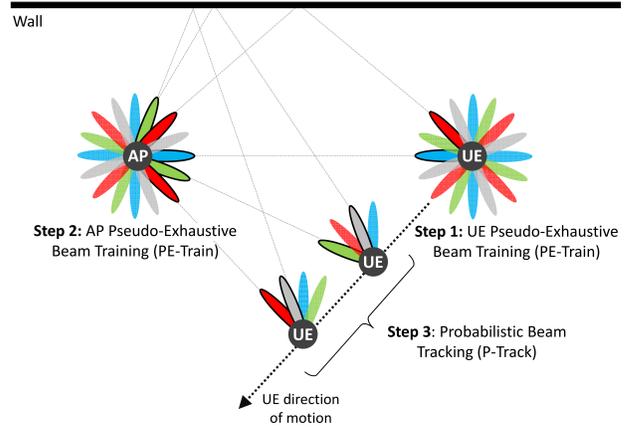

Fig. 4. Typical use case scenario for our beam training strategies. PE-Train is used for initial access beam training and triggered periodically to update the AP-UE steering directions. P-Track, instead, is used to track the mm-wave channel dynamics under node mobility and steer the device beams accordingly.

- The average rate $\overline{R}_\text{D}$ over multiple data frames is obtained by averaging the rate $R_\text{D}$ *over the number of frames* since the latest PE-Train execution.

As for the QoS and timing conditions, we assume that a new PE-Train execution is triggered periodically every $\xi$ pure data frames or when the actual rate $R_\text{D}$ within the current frame is below a certain percentage of the average rate $\overline{R}_\text{D}$, i.e., when $R_\text{D} < \lambda \overline{R}_\text{D}$, with $0 < \lambda \leq 1$.

### B. Ray-tracing module

Geometry-based stochastic channel models in the literature are not suitable to represent real environments, especially time-varying scenarios with user mobility, because they are drop based, meaning that, at every time slot, the channel parameters are randomly generated for each AP-UE link. In order to analyze how the PE-Train and P-Track strategies would perform in a real scenario, we rely in this paper on a custom ray-tracing program in C which allows to deterministically evaluate the non-stationary characteristics of the propagation channel, including LOS and NLOS transitions, shadowing, mobility effects, environment dynamics, and blockage. The key benefit of this approach compared to pure statistical models is its inherent support for spatial consistency which allows smooth and continuous time evolution of channel parameters. As shown in Fig. 3, the ray tracer takes in input the scenario geometry, the electromagnetic characteristics of walls and objects, the AP position, the carrier frequency, and the desired ray-launching resolution (i.e., the angular separation between two adjacent rays launched). The rays are launched to cover the entire 360° azimuthal domain and their evolution is computed taking into account reflections, transmission through objects/walls, and diffraction. At each time slot (i.e., at each UE location), the output from the ray tracer is used to update, in real time during the simulation run, the channel matrix in Eq. 1. To do this, we construct, around the current UE location, a reception sphere with radius proportional to the unfolded path length from AP to UE and the ray-launching resolution [17]. If a ray intersects the sphere, it is taken as contributing to the received signal, otherwise it is discarded. The ray clustering effect revealed by the experimental campaign in [16] is obtained by accounting for the contribution of the ten most powerful rays around the one falling within the reception sphere. This analysis provides the channel AoDs/AoAs at each UE location, while the wave-carried electric field of each ray is used to compute the corresponding complex gain $\alpha_{k\ell}$. Since both phase and delay of each ray are taken into account, the Doppler shift effect is inherently included in the computation.

### C. Simulation scenario

For the performance evaluation, we replicate the 20×20 m² office-like layout considered in [3]. As shown in Fig. 5, it consists of several walls/partitions composed by three different materials, namely concrete, glass, and plasterboard with a thickness of 10 cm, 3 cm, and 5 cm respectively, to model a realistic environment. The scenario geometry and the dielectric properties of materials are given as input to the ray tracer for reflection/transmission/diffraction coefficient calculation. In order to reproduce human blockage effects in a crowded environment, we randomly place in the scenario 20 blocks of size 50×50 cm² with dielectric properties taken from the experiments on human tissues in [18]. The mm-wave network consists of a fixed AP, installed in the center of the room, and a mobile UE walking through three different routes with increasing complexity, namely Route #1 (straight lines with one turn), Route #2 (straight lines with two turns), and Route #3 (curved lines with nine turns). For each route, we assume that the UE is initially located at the starting point (represented by the numbered label in Fig. 5) and is moved with speed $v$=2 m/s and position update rate of $T_\text{slot}$=100 $\mu$s. Note that we consider an orientation-unaware UE, i.e., a UE turn causes the beam orientation to change accordingly.

### D. Results

In this sub-section, we assess the performance of our PE-Train and P-Track strategies using our Matlab/C mm-wave simulator. For performance comparisons, we implement from scratch the solutions proposed in [3] and [5], and a simplified version of the IEEE 802.11ad beam training protocol. An overview of the main characteristics and parameters of the beam search strategies is given in Table I. As for the 802.11ad implementation, we assume that, whenever the QoS falls below a certain threshold, a training frame is allocated to perform a sector level sweep (SLS), while a beam refinement (BRF) procedure for fine grained calibration of the current beams is done every two pure data frames — we verified that this setting provides the best performance for our 802.11ad implementation. We run [5, Algorithm 2], [3, Algorithm 1], and the 802.11ad protocol for different QoS thresholds, and select, for each UE route and for each algorithm, the threshold which provides the best performance. All the simulations consider a 60 GHz carrier frequency with 500 MHz channel bandwidth and a transmit power at both devices equal to 30 dBm. For our PE-Train and P-Track strategies, we assume that 128-length Golay sequences with 4-length Walsh codes are used in both the training slots and the preamble. Concerning the P-Track solution outlined in §V-B, we use 10 gradient descent steps for the preliminary estimation and 50 steps for the final refinement. All the results are averaged over 1000 simulations for each combination of UE route and beam search strategy.

TABLE I
OVERVIEW OF THE MAIN CHARACTERISTICS AND PARAMETERS OF THE BEAM SEARCH STRATEGIES

| Algorithm | Triggering strategy | Transceiver architecture | Algorithm parameters |
|---|---|---|---|
| PE-Train+P-Track | Dynamic QoS and timing thresholds | HBF in Fig. 1 w/ $\lambda/2$-spaced ULAs, $M_{AP}=64$, $N_{AP}=16$, $M_{UE}=24$, $N_{UE}=6$ | 2-bit phase shifters, $L_{est}=2$, $N=1024$, $\overline{\sigma}_{\theta_\ell}=\pi/180$ rad, $f(\text{SNR})=\pi/\text{SNR}$ |
| [5, Algorithm 2] | Fixed QoS threshold | | 7-bit phase shifters, $K=2$, $L_d=L_{est}=2$, $N=1024$ |
| [3, Algorithm 1] | | Unconstrained ABF (phase shifters w/ infinite resolution and amplitude adjustment), | $L_{est}=1$, $\Theta_{AP}=\Theta_{UE}=30°$, $\delta_{AP+UE}=6$ |
| IEEE 802.11ad | | $\lambda/2$-spaced ULAs, $M_{AP}=64$, $M_{UE}=24$ | $L_{est}=1$, SLS and BRF w/ fixed $10°$ beamwidth |

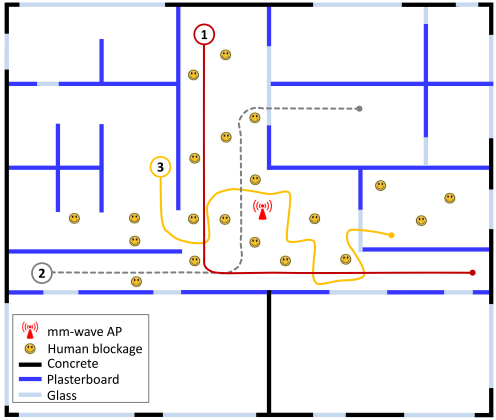

Fig. 5. Office-like simulation scenario with a fixed AP and a mobile UE walking through three different routes.

In Fig. 6, we plot the evolution over time of the achievable normalized rate per frame when different beam search strategies are adopted. We recall that after each beam training/tracking execution, the AP and UE multi-beam antenna patterns are updated according to the new $L_{est}$ estimated steering directions. In the data transmission phase, AP and UE communicate with the narrowest beam patterns they are able to synthesize and transmitting/receiving $N_S = L_{est}$ parallel data streams over $L_{est}$ channel paths. The adopted beam patterns are reflected into the $M_{AP} \times N_S$ data precoder $\mathbf{P}_D$ and the $M_{UE} \times N_S$ data combiner $\mathbf{C}_D$ at the AP and UE respectively. The achievable normalized rate is then calculated as follows:

$$R = \frac{T_{slot}}{T} \sum_{i \in \mathcal{D}} \log_2 \left| \mathbf{I}_{N_S} + \frac{P_t(\mathbf{U}^H \mathbf{H}_i \mathbf{P}_D)(\mathbf{U}^H \mathbf{H}_i \mathbf{P}_D)^H}{N_S \sigma^2} \right| \quad (11)$$

where $\mathbf{I}_{N_S}$ is the $N_S \times N_S$ identity matrix, $\mathcal{D}$ is the set of data slots in the frame, $\mathbf{H}_i$ is the channel matrix in the $i$-th slot, $\sigma^2$ is the average noise power, and $\mathbf{U}$ is left singular vector matrix of the "economic" SVD decomposition of $\mathbf{C}_D$.

As evident from the plots, our simulator is able to well describe the effect of human blockage at mm-wave frequencies. The blockage, which intermittently appears and breaks the LOS link between AP and UE, is clearly visible from the rate suddenly dropping down to very small values. In case even the optimum rate sharply drops to zero, there is no possibility to establish a connection between AP and UE, i.e., the UE is in outage. In all the analyzed UE routes, our strategy, based on the alternation of PE-Train and P-Track according to the selected QoS and timing thresholds, yields performance very close to the optimum oracle algorithm (only 10% rate difference on average), and significantly outperforms both the 802.11ad approach and beam search proposals in the literature. Quantitatively, based on the results in Fig. 6, we report in Table II the average training overhead $\overline{\tau}$ per frame, calculated by dividing the total number of training slots used in each route by the number of allocated frames and multiplying by the time slot duration $T_{slot}$. As shown, our PE-Train and P-Track strategies provide a one to two orders of magnitude reduction in training overhead. This translates, approximately, to an average rate increase of 48% to 150% compared to state-of-the-art solutions and of 40% to 50% over the 802.11ad standard. Although these results are obtained for $\lambda=0.3$ and $\xi=30$ frames, we verified via grid search that the performance achieved by our approach does not vary significantly with the selected QoS and timing thresholds. In fact, both the continuous execution of P-Track and the low-overhead PE-Train make the choice of $\lambda$ and $\xi$ not determinant on the overall performance. We repeated the evaluation for a vast range of simulation scenarios and UE routes (not reported here due to space constraints), where we noted that the achieved performance does not differ considerably from that presented here. We also verified that our strategies perform well in environments with less blockage. For example, in the scenario of Fig. 5 without human blockage, we obtained a 25% to 170% rate increase over existing approaches. It is worth emphasizing that, differently from [5] and 802.11ad, our strategies do not require any dedicated channel for the receiver to feed back the training results to the transmitter. Furthermore, we adopt HBF with only 2-bit phase shifters as opposed to the 7-bit ones used in [5] and the idealized, unconstrained ABF transceiver considered for the implementation of [3] and 802.11ad.

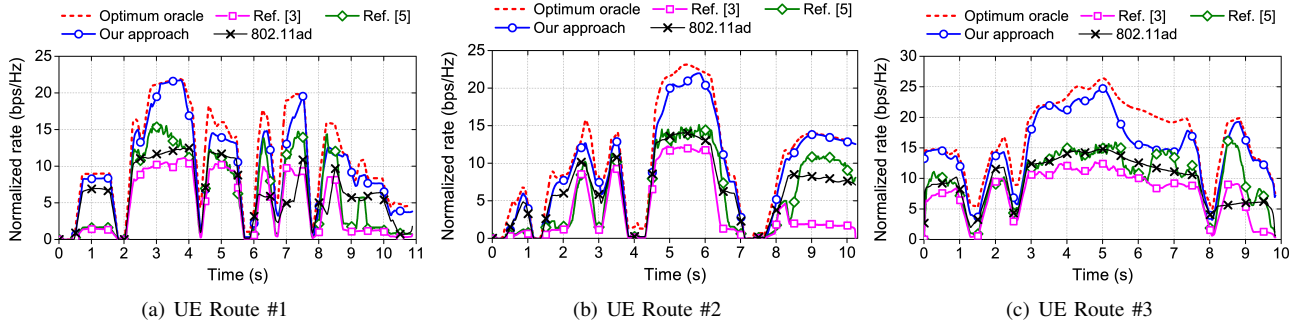

(a) UE Route #1     (b) UE Route #2     (c) UE Route #3

Fig. 6. Normalized rate over time for the three UE routes: comparison among the optimum oracle solution, the proposed strategies with $\lambda$=0.3 and $\xi$=30 frames), two beam training algorithms in the literature, and the baseline IEEE 802.11ad protocol.

TABLE II
TRAINING OVERHEAD AND RATE GAIN PROVIDED BY OUR APPROACH

| Algorithm | Training overhead $\bar{\tau}$ (ms) | | | Percentage rate gain of PE-Train+P-Track | | |
|---|---|---|---|---|---|---|
| | R#1 | R#2 | R#3 | R#1 | R#2 | R#3 |
| PE-Train+P-Track | 0.06 | 0.04 | 0.06 | – | – | – |
| [5, Algorithm 2] | 4.62 | 4.35 | 3.07 | 56.1 | 54.8 | 47.7 |
| [3, Algorithm 1] | 3.58 | 4.67 | 1.66 | 108.2 | 150.1 | 94.1 |
| IEEE 802.11ad | 0.87 | 0.72 | 0.67 | 44.0 | 39.8 | 50.2 |

## VII. CONCLUSION

In this paper, we investigated the problem of beam training and tracking in directional mm-wave networks with mobility. Exploiting the ability of HBF transceivers to collect channel information from multiple spatial directions simultaneously, we designed two strategies (one deterministic for beam training and one probabilistic for beam tracking) to rapidly estimate the most suitable transmit/receive directions at the AP and UE sides. Simulation results, obtained by a custom simulator based on ray-tracing channel modeling, demonstrated that the proposed solution is effective to keep the average communication rate only 10% below the optimal bound. Compared to both the IEEE 802.11ad standard and the state of the art, our solution provides a 40% to 150% performance increase while at the same time using lower complexity hardware.

## APPENDIX A
### DERIVATION OF THE MEASUREMENT OBJECTIVE FUNCTION

In this appendix, we derive the expression of $O_Y(\boldsymbol{\theta})$ given in Eq. 10. For the sake of brevity, we only refer to the estimation of the most promising transmit/receive directions performed by the UE through the processing of downlink data slots. Similar considerations hold also for the estimation at the AP side through uplink data slots. In the probabilistic optimization problem formulated in Eq. 8, differently from $O_P(\boldsymbol{\theta})$, the measurement objective function $O_Y(\boldsymbol{\theta}) = -\log[P(\mathbf{Y}_{\text{RF}}|\boldsymbol{\theta})]$ models those changes in the mm-wave channel (due to the user mobility) which are directly reflected into the received signal at the UE. We assume that AP and UE are communicating using $L_{\text{est}}$ parallel streams of data. Each data stream is transmitted/received using the narrowest beams the devices are able to synthesize, steered towards the most powerful directions estimated in the most recent PE-Train/P-Track execution. We can approximate the mm-wave channel with $L_{\text{est}}$ dominant paths as:

$$\hat{\mathbf{H}} = \sqrt{M_{\text{AP}} M_{\text{UE}}} \sum_{\ell=1}^{L_{\text{est}}} \alpha_\ell \mathbf{a}_{\text{UE}}(\theta_\ell) \mathbf{a}_{\text{AP}}^H(\phi_\ell) \quad (12)$$

where $\theta_\ell$ and $\phi_\ell$ are AoDs/AoAs (i.e., the steering directions) exploited by UE and AP respectively, and $\alpha_\ell$ is the relative complex gain. Since the received signal at the UE is affected by AWGN with power $\sigma^2$, the probability of measuring $\hat{\mathbf{Y}}_{\text{RF}}$ conditioned to the channel characteristics is given by:

$$P(\hat{\mathbf{Y}}_{\text{RF}}|\boldsymbol{\theta}, \boldsymbol{\phi}, \boldsymbol{\alpha}) = K_\sigma e^{-\frac{\|\mathbf{D}^{-1}\mathbf{V}^H \hat{\mathbf{Y}}_{\text{RF}} - \mathbf{U}^H \hat{\mathbf{H}} \mathbf{P}\|_F^2}{2\sigma^2}} \quad (13)$$

where $\mathbf{UDV}^H$ is the economic SVD of the UE RF precoder $\mathbf{C}_{\text{RF}}$, $\mathbf{P}$ is the AP hybrid precoder, $K_\sigma$ is the normalization factor, and $\boldsymbol{\theta}$, $\boldsymbol{\phi}$, and $\boldsymbol{\alpha}$ are the $L_{\text{est}} \times 1$ vectors containing respectively the AoD/AoAs at the UE, the AoDs/AoAs at the AP, and the respective complex gains. We handle the lack of knowledge about the AP hybrid precoder $\mathbf{P}$ by incorporating it into the channel and defining:

$$\hat{\mathbf{H}}\mathbf{P} = \mathbf{A}_\theta \mathbf{M} \quad (14)$$

where $\mathbf{A}_\theta$ is a $M_{\text{UE}} \times L_{\text{est}}$ matrix such that $[\mathbf{A}_\theta]_{:,\ell} = \mathbf{a}_{\text{UE}}(\theta_\ell)$ and $\mathbf{M}$ is a $L_{\text{est}} \times L_{\text{est}}$ matrix such that $[\mathbf{M}]_{\ell,:} = \alpha_\ell \mathbf{a}_{\text{AP}}^H(\theta_\ell)^H \mathbf{P}$. Since we are interested in an objective function which depends only on the UE steering directions $\boldsymbol{\theta}$, we suppose that $\mathbf{M}$ is unconstrained, i.e., we ignore $\mathbf{P}$ and $\boldsymbol{\alpha}$, and define the relaxed measurement objective function by taking the negative loga-rithm of the expression in Eq. 13 and removing unnecessary constant terms:

$$\hat{O}_Y(\boldsymbol{\theta}, \mathbf{M}) = \frac{\|\mathbf{D}^{-1}\mathbf{V}^H\hat{\mathbf{Y}}_{\text{RF}} - \mathbf{U}^H\mathbf{A}_\theta\mathbf{M}\|_F^2}{2\sigma^2} \quad (15)$$

In order to remove the dependance on $\mathbf{M}$, we finally define the measurement objective function as

$$O_Y(\boldsymbol{\theta}) = \arg\min_{\mathbf{M}} \hat{O}_Y(\boldsymbol{\theta}, \mathbf{M}) \quad (16)$$

which can be easily computed, since the matrix $\mathbf{M}$ satisfying Eq. 16 is given by the solution of a minimum mean square error (MMSE) problem:

$$\mathbf{M} = (\mathbf{A}_\theta^H \mathbf{U}\mathbf{U}^H \mathbf{A}_\theta)^{-1} \mathbf{A}_\theta^H \mathbf{U}\mathbf{D}^{-1}\mathbf{V}^H\hat{\mathbf{Y}}_{\text{RF}} \quad (17)$$

which leads to the final expression of the measurement objective function:

$$O_Y(\boldsymbol{\theta}) = \frac{\|\mathbf{D}^{-1}\mathbf{V}^H\hat{\mathbf{Y}}_{\text{RF}} - \mathbf{U}^H\mathbf{A}_\theta\mathbf{M}\|_F^2}{2\sigma^2} \quad (18)$$

It is easy to see that a good initial guess $\boldsymbol{\theta}$ for $O_Y(\boldsymbol{\theta})$ is the set of directions with maximum received signal power at the UE. Such an initial guess can be used as a starting point for the solution of the optimization problem in §V-B and can be easily obtained by finding the discrete angle $\theta_\ell$, with $\ell = 1, 2, ..., L_{\text{est}}$, which minimizes the following expression:

$$\left\|\left(\mathbf{D}^{-1}\mathbf{V}^H - \mathbf{U}^H\mathbf{a}_{\text{UE}}(\theta_\ell)\frac{\mathbf{a}_{\text{UE}}(\theta_\ell)^H\mathbf{U}\mathbf{D}^{-1}\mathbf{V}^H}{\mathbf{a}_{\text{UE}}(\theta_\ell)^H\mathbf{U}\mathbf{U}^H\mathbf{a}_{\text{UE}}(\theta_\ell)}\right)[\hat{\mathbf{Y}}_{\text{RF}}]_{:,\ell}\right\|_F^2 \quad (19)$$

## APPENDIX B
### GRADIENTS OF THE OBJECTIVE FUNCTIONS

In this appendix, we derive the expressions for the gradient of the prior objective function $O_P(\boldsymbol{\theta})$ and the measurement objective function $O_Y(\boldsymbol{\theta})$, which are required to solve the optimization problem in Eq. 8 through the gradient descent approach outlined in §V-B. For the sake of brevity, we only derive the expressions referred to the estimation performed at the UE side. The extension to the AP side is straightforward.

As for the prior objective function $O_P(\boldsymbol{\theta})$ in Eq. 9, the gradient vector entries can be computed as:

$$\nabla_\ell O_P(\boldsymbol{\theta}) = \frac{\theta_\ell - \overline{\theta}_\ell}{\sigma_{\theta_\ell}^2}, \quad \ell = 1, 2, ..., L_{\text{est}} \quad (20)$$

As for the measurement objective function $O_Y(\boldsymbol{\theta})$ in Eq. 10, we start by manipulating its expression as follows:

$$O_Y(\boldsymbol{\theta}) = \frac{\|\Phi\|_F^2}{2\sigma^2} = \frac{\sum_{\ell=1}^{L_{\text{est}}}\|[\Phi]_{:,\ell}\|^2}{2\sigma^2} \quad (21)$$

where $\Phi = \mathbf{D}^{-1}\mathbf{V}^H\hat{\mathbf{Y}}_{\text{RF}} - \mathbf{U}^H\mathbf{A}_\theta\mathbf{M}$. Applying the gradient operator to the previous expression and, then, the chain rule, we obtain:

$$\nabla_i O_Y(\boldsymbol{\theta}) = \frac{\sum_{\ell=1}^{L_{\text{est}}} \text{Re}\left\{[\Phi]_{:,\ell}^H \nabla_\ell [\Phi]_{:,\ell}\right\}}{\sigma^2} \quad (22)$$

for $i = 1, 2, ..., L_{\text{est}}$. The term $\nabla_i[\Phi]_{:,\ell}$ can be expressed as:

$$\nabla_i[\Phi]_{:,\ell} = -\mathbf{U}^H(\nabla_i\mathbf{A}_\theta)[\mathbf{M}]_{:,\ell} - \mathbf{U}^H\mathbf{A}_\theta(\nabla_i[\mathbf{M}]_{:,\ell}) \quad (23)$$

Applying the orthogonality principle to the definition of $\mathbf{M}$, we have that, in the expression of $[\Phi]_{:,\ell}^H \nabla_\ell [\Phi]_{:,\ell}$ in Eq. 22, the second term given by $-\text{Re}\{[\Phi]_{:,\ell}^H \mathbf{U}^H \mathbf{A}_\theta (\nabla_i[\mathbf{M}]_{:,\ell})\}$ is canceled, leading to:

$$\text{Re}\left\{[\Phi]_{:,\ell}^H \nabla_\ell [\Phi]_{:,\ell}\right\} = -\text{Re}\left\{[\Phi]_{:,\ell}^H \mathbf{U}^H (\nabla_i \mathbf{A}_\theta)[\mathbf{M}]_{:,\ell}\right\} \quad (24)$$

The term $\nabla_i \mathbf{A}_\theta$ can be computed as:

$$\nabla_i \mathbf{A}_\theta = -j(\mathbf{A}_{\text{ind}} \circ \mathbf{A}_\theta)\mathbf{E}_i \quad (25)$$

where $j$ is the imaginary unit, $\mathbf{A}_{\text{ind}}$ is a $M_{\text{UE}} \times L_{\text{est}}$ matrix such that $[\mathbf{A}_{\text{ind}}]_{n,m} = n - 1$, $\mathbf{E}_i$ is a $L_{\text{est}} \times L_{\text{est}}$ matrix with all zeros except a one at position $[\mathbf{E}_i]_{i,i}$, and $\circ$ denotes the element-wise multiplication (Hadamard product). Substituting Eq. 25 into Eq. 24, we obtain:

$$\text{Re}\left\{[\Phi]_{:,\ell}^H \nabla_\ell [\Phi]_{:,\ell}\right\} = -\text{Im}\left\{[\Phi]_{:,\ell}^H \mathbf{U}^H (\mathbf{A}_{\text{ind}} \circ \mathbf{A}_\theta)\mathbf{E}_i[\mathbf{M}]_{:,\ell}\right\} \quad (26)$$

Then, substituting Eq. 26 in Eq. 22, we finally obtain:

$$\nabla_i O_Y(\boldsymbol{\theta}) = -\frac{\sum_{\ell=1}^{L_{\text{est}}} \text{Im}\left\{[\Phi]_{:,\ell}^H \mathbf{U}^H (\mathbf{A}_{\text{ind}} \circ \mathbf{A}_\theta)\mathbf{E}_i[\mathbf{M}]_{:,\ell}\right\}}{\sigma^2} \quad (27)$$

which, by concatenation, leads to the final expression for the gradient of the measurement objective function:

$$\nabla O_Y(\boldsymbol{\theta}) = -\frac{\sum_{\ell=1}^{L_{\text{est}}} \text{Im}\left\{\left[[\Phi]_{:,\ell}^H \mathbf{U}^H (\mathbf{A}_{\text{ind}} \circ \mathbf{A}_\theta)\right]^T \circ [\mathbf{M}]_{:,\ell}\right\}}{\sigma^2} \quad (28)$$

ACKNOWLEDGMENTS

The research leading to these results received funding from the European Commission H2020 programme under grant agreement n◦ 671650 (5G PPP mmMAGIC project). This article was also partially supported by the Madrid Regional Government through the TIGRE5-CM program (S2013/ICE-2919), the Ramon y Cajal grant from the Spanish Ministry of Economy and Competitiveness RYC-2012-10788, and the European Research Council grant ERC CoG 617721.